\documentclass[12pt,journal,draftclsnofoot,onecolumn]{IEEEtran}
\usepackage{amsfonts}
\usepackage{amssymb}
\usepackage{cite}
\usepackage[cmex10]{amsmath}
\usepackage{float}
\usepackage{color}
\usepackage{stfloats,fancyhdr}
\usepackage{amsmath}
\usepackage{amsthm}
\usepackage{enumerate}
\usepackage{algorithm}
\usepackage{algorithmic}
\usepackage{multirow}
\usepackage{changepage}
\usepackage[normalem]{ulem}

\usepackage{url}
\usepackage{xr-hyper}
\usepackage{hyperref}

\makeatletter
\newcommand*{\addFileDependency}[1]{
  \typeout{(#1)}
  \@addtofilelist{#1}
  \IfFileExists{#1}{}{\typeout{No file #1.}}
}
\makeatother

\IEEEoverridecommandlockouts

\ifCLASSINFOpdf
  \usepackage[pdftex]{graphicx}
  \DeclareGraphicsExtensions{.pdf,.jpeg,.png}
\else
  \usepackage[dvips]{graphicx}
  \DeclareGraphicsExtensions{.pdf}
\fi

\usepackage{subfigure}

\usepackage{fancybox,dashbox}

\begin{document}
%
\title{Is Multichannel Access Useful in Timely Information Update?}

\author{Jiaxin~Liang,
        Haoyuan~Pan,~\IEEEmembership{Member,~IEEE,}
        and~Soung~Chang~Liew,~\IEEEmembership{Fellow,~IEEE}%
\thanks{J. Liang, and S. C. Liew are with Department of Information Engineering, The Chinese University of Hong Kong, Hong Kong SAR, China (email: \{lj015, soung\}@ie.cuhk.edu.hk).}
\thanks{H. Pan is with the College of Computer Science and Software Engineering, Shenzhen University, Shenzhen, 518060, China (email: hypan@szu.edu.cn).}}

\maketitle

\begin{abstract}
This paper investigates information freshness of multichannel access in information update systems. Age of information (AoI) is a fundamentally important metric to characterize information freshness, defined as the time elapsed since the generation of the last successfully received update. When multiple devices share the same wireless channel to send updates to a common receiver, an interesting question is whether dividing the whole channel into several subchannels will lead to better AoI performance. Given the same frequency band, dividing it into different numbers of subchannels lead to different transmission times and packet error rates (PER) of short update packets, thus affecting information freshness. We focus on a multichannel access system where different devices take turns to transmit with a cyclic schedule repeated over time. We first derive the average AoI by estimating the PERs of short packets. Then we examine bounded AoI, for which the instantaneous AoI is required to be below a threshold a large percentage of the time. Simulation results indicate that multichannel access can provide low average AoI and uniform bounded AoI simultaneously across different received powers. Overall, our investigations provide insights into practical designs of multichannel access systems with AoI requirements.
\end{abstract}



\IEEEpeerreviewmaketitle

\section{Introduction}
Next-generation wireless networks are expected to support time-sensitive Internet of Things (IoT) services and applications, such as motion control and environmental quality monitoring in industrial IoT \cite{iotsurvey}. Timely delivery and update of information are critical to these systems \cite{aoimagazine}.

The design of conventional communication networks focuses on delays and throughput as the performance metrics. A new performance metric, age of information (AoI), has been introduced to quantify information freshness at the receiver \cite{Kosta_AoI_book}. AoI captures both the packet generation time and packet delay. Specifically, AoI is the age of the information last received by the receiver---if at time $t$, the latest sample received by the receiver was a sample generated at the source at time $t'$, then the instantaneous AoI at time $t$ is $t - t'$. Prior works show that replacing throughput and delay by AoI as the performance metric may lead to fundamentally different communication system designs.  \cite{aoimagazine,Kosta_AoI_book,haoyuan_aoi,aoi_worstcase,aoi_queue,WaitingBeforeServing, aoi_schedule}.

In many IoT information update systems, multiple devices send the statuses of the remote processes that they monitor via a shared wireless channel to a common receiver \cite{haoyuan_aoi,aoi_worstcase} . Given the same frequency band, we could divide the wireless channel into several frequency subchannels for wireless access. We refer to this system as a multichannel access system. When the number of subchannels is large, more devices can access the wireless channel at the same time, leading to possible AoI drops of multiple devices simultaneously. In addition, more subchannels mean that smaller bandwidths are allocated to devices. As a result, devices with fixed transmit power, when transmitting signals over the narrowband, can have lower PER. However, smaller bandwidths lead to longer transmission time of packets as well as a possibly higher AoI, because it would take a longer time to transmit the same amount of information and consequently the information received was generated at a time further in the past. A quantitative study is needed to evaluate the relative merits of using different numbers of subchannels when AoI is the performance metric.

Although AoI has attracted considerable research interests, prior works focuses on queuing models and packet scheduling policies \cite{Kosta_AoI_book,aoi_queue,WaitingBeforeServing, aoi_schedule} . How to utilize precious time-frequency resources has not been well investigated. For example, \cite{haoyuan_aoi} focused on different devices using the whole bandwidth to transmit. Although different frequency bands are also considered in \cite{haoyuan_aoi}, the authors assumed that the number of frequency bands is equal to the number of devices, which is impractical with massive IoT devices. 

\begin{figure}
\centering
\includegraphics[width=0.8\textwidth]{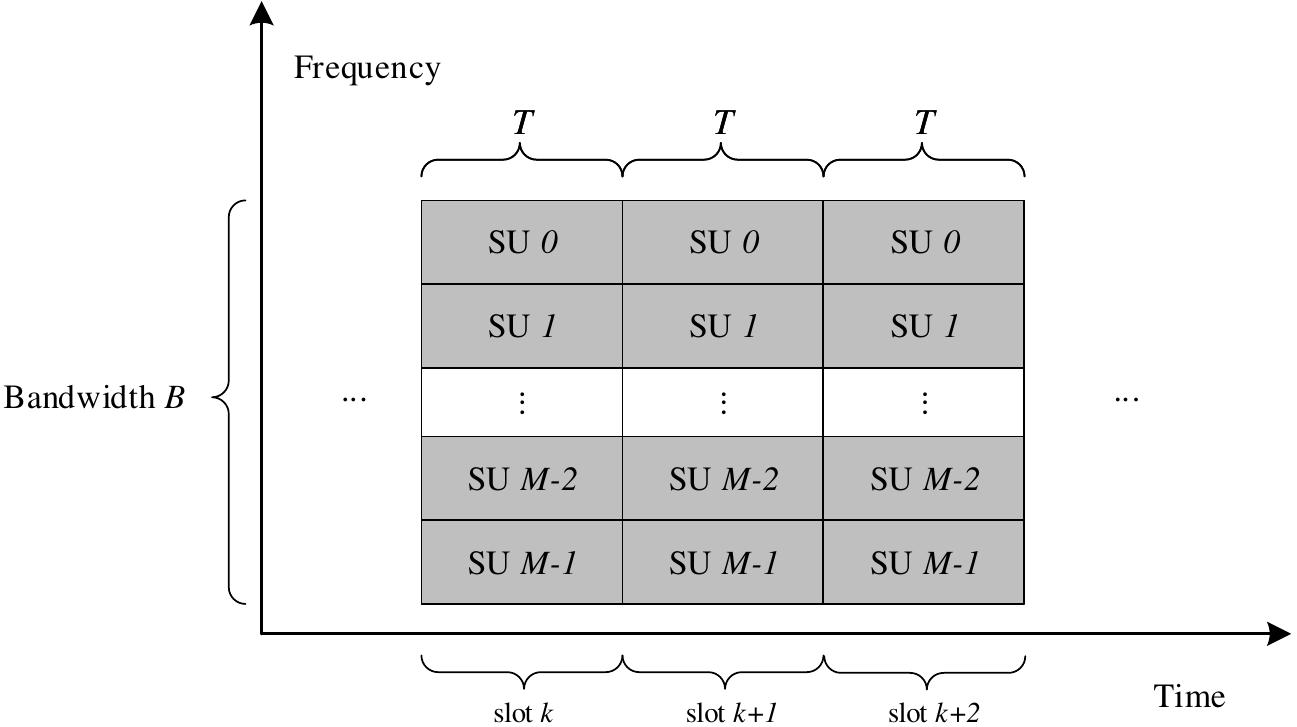}
\caption{The channel resource is divided in both the frequency domain and time domain for the multichannel access systems. The gray blocks are basic units for scheduling. They are referred to as scheduling units (SU).}\label{fig:resource_map}
\vspace{-0.2in}
\end{figure}

To efficiently exploit the available bandwidth, we investigate the AoI performance of a multichannel access system with $M$ independent parallel frequency bands, as shown in Fig. \ref{fig:resource_map}. Time is divided into time slots and the basic time-frequency block is called a scheduling unit (SU). An SU is assigned to a single link between a device and the receiver. We consider a generic communication schedule that is cyclically repeated over time so that every device has an update opportunity in each cycle (the details can be found in Section \ref{sec:aoi_multichannel}). Since packets are typically short in information update systems, we use the Polyanskiy-Poor-Verdu bound \cite{per_bound} to estimate the packet error rate (PER) of packets over unreliable channels.

We considers two AoI metrics, average AoI and bounded AoI. Average AoI measures the instantaneous AoI averaged over time. However, it is not sufficient to have low average AoI in many time-critical systems when the instantaneous AoI also needs to be upper bounded by a predefined threshold. Bounded AoI \cite{haoyuan_aoi} is an AoI threshold that the instantaneous AoI does not exceed for a target percentage of the time $\epsilon$. Specifically, the percentage of time the instantaneous AoI is below the threshold must be larger than or equal to $\epsilon$. Given a target $\epsilon$, a lower bounded AoI means that the update system can provide a higher level of information freshness.

Simulation results show that when the received power is medium or low, multichannel access can outperform both TDMA and FDMA in terms of average AoI and bounded AoI. Moreover, multichannel access can effectively balance between average AoI and bounded AoI performance. Specifically, by adjusting the number of subchannels, multichannel access can maintain low average AoI and uniform bounded AoI simultaneously across different received powers, availing the system of robust operation when the received powers from the devices are non-uniform. 

\section{Backgrounds}
\subsection{Age of Information (AoI) Metrics}


We study an information update system with $N$ IoT devices and an access point (AP). We denote the time the $i$-th device $u_i$ sends the $j$-th information update packet by $t_j^i$. The packet is received by the AP at time $t_j^{i'}$. At any given time $t$, the last update packet from device $i$ received by the AP is denoted by ${N_i}(t) = \max \{ j|t_j^{i'} \le t\}$. Let ${U_i}(t) = {t_{{N_{\rm{i}}}(t)}}$ be the time that packet was generated at device $i$. The \textit{instantaneous AoI} of device $i$ at time $t$ is defined by
\begin{equation}	\label{eqa:inst_aoi}
{\Delta _i}(t) = t - {U_i}(t).
\end{equation}
${\Delta _i}(t)$ is commonly used as an intermediary to compute other AoI metrics \cite{Kosta_AoI_book}. This paper focuses on two AoI metrics, namely average AoI and bounded AoI, as presented below.

\subsubsection*{Average AoI}
Average AoI is commonly evaluated in information update systems. Specifically, it is the time average of the instantaneous AoI $\Delta _i$ \cite{Kosta_AoI_book} given by
\begin{equation}	\label{eqa:avg_aoi}
{\bar \Delta _i} = \mathop {\lim }\limits_{T \to \infty } \frac{1}{T}\int_0^T {{\Delta _i}(t)dt}.
\end{equation}
The average AoI of all devices in the system is $\bar \Delta  = \frac{1}{N}\sum\limits_{i = 1}^N {{{\bar \Delta }_i}}$.

\subsubsection*{Bounded AoI}
In practice, it is not sufficient to have a low average AoI for many information update systems. Bounded AoI was first proposed in \cite{haoyuan_aoi} for time-critical systems that require the instantaneous AoI ${\Delta _i}(t)$ to be upper bounded by a predefined threshold $\Delta _{THR}$. The threshold $\Delta _{THR}$ is determined by the system's timing requirement: the percentage of time the instantaneous AoI is below $\Delta _{THR}$ is larger than or equal to a target value $\epsilon$. Given a target $\epsilon$, we define bounded AoI   as the smallest value that satisfies 
\begin{equation}	\label{eqa:bound}
{O_{i,\epsilon}} = \mathop {\lim }\limits_{T \to \infty } \frac{1}{T}\int_0^T {I[{\Delta _{THR}} \ge {\Delta _i}(t)]dt}  \ge \epsilon,\forall i \in \{ 1,...,N\},
\end{equation}
where $I[{\Delta _{THR}} \ge {\Delta _i}(t)]$ is an indicator function, which is equal to 1 if $\Delta _{THR} \ge {\Delta _i}(t)$, and 0 otherwise. The metric $O_{i,\epsilon}$ corresponds to the fraction of time that ${\Delta _i}(t)$ is bounded by $\Delta _{THR}$. If (\ref{eqa:bound}) is satisfied, we say that a system can provide a bounded AoI $\Delta _{THR}$ with confidence $\epsilon$. It is easy to see that a lower $\Delta _{THR}$ means that the system can provide fresher information with a certain confidence.


We can estimate $\Delta _{THR}^{}$ by Chebyshev's inequality, which characterizes the behaviors of bounded AoI well \cite{haoyuan_aoi}. Let $\Delta _{THR[i]}$ denote the bounded AoI for device $i$. We have
\begin{align}	
\label{eqa:bound_derive}
{O_{i,\epsilon}} &= \mathop {\lim }\limits_{T \to \infty } \frac{1}{T}\int_0^T {I[{\Delta _{THR}} \ge {\Delta _i}(t)]dt} \notag \\
&\mathop  = \limits^{(1)} 1 - \mathop {\lim }\limits_{T \to \infty } \frac{1}{T}\int_0^T {I[({\Delta _i}(t) - {{\bar \Delta }_i}) \ge ({\Delta _{THR[i]}} - {{\bar \Delta }_i})]dt} \notag \\
&\mathop  \ge \limits^{(2)} 1 - \mathop {\lim }\limits_{T \to \infty } \frac{1}{T}\int_0^T {I[|{\Delta _i}(t) - {{\bar \Delta }_i}| \ge |{\Delta _{THR[i]}} - {{\bar \Delta }_{\rm{i}}}|]dt} \notag\\
&\mathop  \ge \limits^{(3)} 1 - {\rm{ }}\frac{{\sigma _{\rm{i}}^2}}{{{{({\Delta _{THR[i]}} - {{\bar \Delta }_i})}^2}}},
\end{align}

\noindent where the inequality (3) in (\ref{eqa:bound_derive}) is obtained by Chebyshev's inequality. Define $\sigma _{\rm{i}}^2$ as the variance of the instantaneous AoI over time of device $i$
\begin{equation}	\label{eqa:bound_variance}
\sigma _{\rm{i}}^2 = \mathop {\lim }\limits_{T \to \infty } \frac{1}{T}\int_0^T {{{\left( {{\Delta _i}(t) - {{\bar \Delta }_i}} \right)}^2}} dt{\rm{ = }}\overline {\Delta _i^2}  - {\left( {{{\bar \Delta }_i}} \right)^2},
\end{equation}
where $\overline {\Delta _{\rm{i}}^2}  = \mathop {\lim }\limits_{T \to \infty } \frac{1}{T}\int_0^T {{\Delta _i}{{(t)}^2}} dt$. Given a fixed $\epsilon$, we compute an upper bound ${\hat \Delta _{THR[{\rm{i}}]}}$ for the corresponding bounded AoI $\Delta _{THR[i]}$ for device $i$ by
\begin{equation}	\label{eqa:upper_bound}
{\hat \Delta _{THR[{\rm{i}}]}} = \sqrt {\frac{{\sigma _i^2}}{{1 - \epsilon}}}  + {\bar \Delta _i} = \sqrt {\frac{{\overline {\Delta _i^2}  - {{\left( {{{\bar \Delta }_i}} \right)}^2}}}{{1 - \epsilon}}}  + {\bar \Delta _i}.
\end{equation}
Finally, when all the devices in a network are considered, we have an upper bound ${\hat \Delta _{THR}}$ for the network's bounded AoI
\begin{equation}	\label{eqa:network_bound_aoi}
{\hat \Delta _{THR}} = {\max _{i \in \{ 1,2,...,n\} }}{\hat \Delta _{THR[{\rm{i}}]}}.
\end{equation}

\subsection{Packet Error Rates for Short Packets}
In practical information update systems, update packets are typically short. With finite block lengths, the packet error rate (PER) cannot go to zero. According to \cite{per_bound}, the PER $p$ in AWGN channels can be approximated as
\begin{equation}	\label{eqa:per_bound}
p \approx Q\left( {\frac{{\frac{1}{2}{{\log }_2}(1 + \gamma ) - \frac{K}{L}}}{{{{\log }_2}(e)\sqrt {\frac{1}{{2L}}\left( {1 - \frac{1}{{{{\left( {1 + \gamma } \right)}^2}}}} \right)} }}} \right),
\end{equation}
where $K$ is the number of source bits of a packet; $L$ is the block length of coded bits; $Q( \cdot )$ is the Q-function. $\gamma $ is the effective signal-to-noise ratio (SNR) at the receiver. We remark that the AoI analysis provided in this paper can be readily extended to fading channels by replacing the expression of PER in (\ref{eqa:per_bound}) with those of fading channels derived in \cite{yifan_2018_short}.

\section{AoI in Multichannel Access Systems}	\label{sec:aoi_multichannel}
This section analyses the theoretical AoI performance of the multichannel access system. Section \ref{subsec:system_model} first presents the system model. We consider symmetric channels in which all devices have the same received power at the receiver. Section \ref{subsec:aoi_in_multichannel} then derives the average AoI and bounded AoI.

\subsection{System model}	\label{subsec:system_model}
\begin{figure}
\centering
\includegraphics[width=0.8\textwidth]{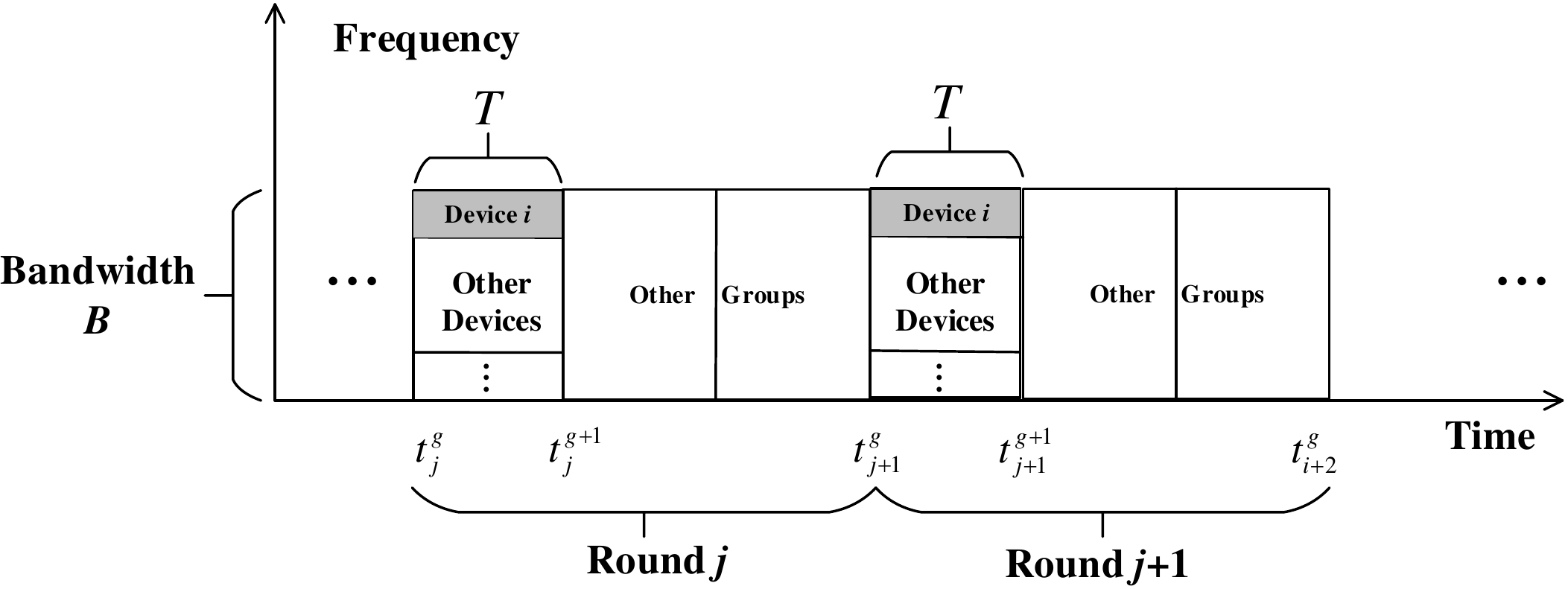}
\caption{Multichannel access system model. Devices are divided into groups and each group will take turns to access the channel.}
\label{fig:schedule_scheme}
\vspace{-0.2in}
\end{figure}

In a multichannel access system as shown in Fig. \ref{fig:schedule_scheme}, the $N$ devices are divided into $G=N/M$ groups, where $M$ the number of devices per group. In the time domain, $G$ consecutive time slots are defined as one round, in which different time slots are then allocated to the $G$ different groups. Each group occupies one time slot. Within a group, the frequency band is divided into $M$ non-overlapping channels, and each device is then assigned a different channel for multiple access. That is, the $M$ devices in the same group send at the same time but in different frequency bands. As special cases, $M = 1$ for pure TDMA systems and $M = N$ for pure FDMA systems.

The communication schedule is cyclically repeated over time. Suppose that different groups send update packets to the receiver in a round-robin manner. We define a round as the total transmission time of all the $G$ groups. Since with symmetric channels all devices have the same received power $P$, we assume each group has the same time slot duration $T$, and the devices in the same group are allocated with equal bandwidth $B/M$. The time duration of a round is then $GT$.

In practical information update systems, physical observations of the IoT devices on remote processes are made and submitted at regular intervals. Suppose that in every round, each device generates a new update packet. Let interval $[{t_j},{t_{j + 1}}]$ denote the time period of round $j$, i.e., ${t_{j + 1}} = {t_j} + GT$. In round $j$, device $i$ in the $g$-th group generates packet $C_j^i$ at time $t_j^g$ and finishes transmission at time $t_j^{g + 1} = t_j^g + T$, where $t_j^g = {t_j} + (g - 1)T$. In other words, the devices in the same group send packets at an interval $GT$.

\begin{figure}
\centering
\includegraphics[width=0.8\textwidth]{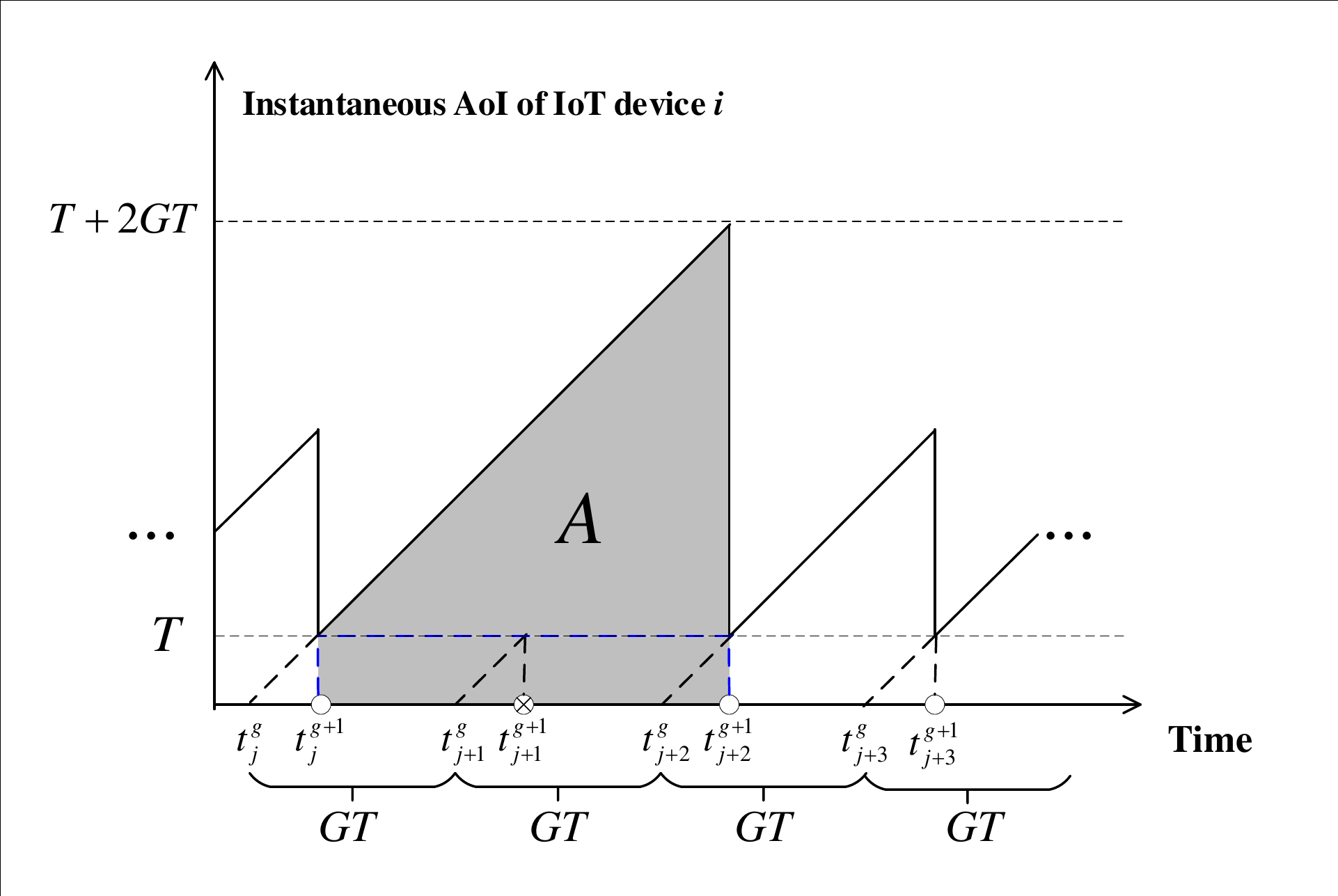}
\caption{An example of ${\Delta _i}(t)$ in an $N$-device multichannel access system. Three update packets are successfully decoded at $t_j^{g + 1}$, $t_{j+2}^{g + 1}$, and $t_{j+3}^{g + 1}$, while one packet fails to be decoded at $t_{j + 1}^{g + 1}$.}\label{fig:aoi_example}
\vspace{-0.2in}
\end{figure}

In round $j$, packet decoding for group $g$ occurs at $t_j^{g + 1}$. If device $i$'s packet $C_j^i$ is successfully decoded by the receiver, the instantaneous AoI ${\Delta _i}(t)$ is reset to $T$. Fig. \ref{fig:aoi_example} shows an example of ${\Delta _i}(t)$, where ${\Delta _i}(t)$ drops to $T$ at times $t_j^{g + 1}$, $t_{j+2}^{g + 1}$, and $t_{j+3}^{g + 1}$ because of successful packet decoding. One packet fails to be decoded at time $t_{j + 1}^{g + 1}$, so ${\Delta _i}(t)$ continues to increase linearly. 

\subsection{Derivations of Average AoI and Bounded AoI}	\label{subsec:aoi_in_multichannel}
\subsubsection*{Average AoI $\bar \Delta$} Suppose that device $i$'s packet $C_j^i$ is successfully decoded with probability $1 - p$, where $p \in [0,1]$ denotes the PER. Denote by $X$ the number of packets transmitted between two successful packet decodings. Then $X$ is a geometric random variable with a probability mass function ${P_X}(X = x) = {(p)^{x - 1}}(1 - p),x = 1,2,...$, i.e., $x - 1$ failed decodings followed by one successful decoding\footnote{Note that $E[X] = \frac{1}{{1 - p}}$, $E[{X^2}] = \frac{{1 + p}}{{{{(1 - p)}^2}}}$, and $E[{X^3}] = 1 + \frac{{7p}}{{1 - p}} - \frac{{12{p^2}}}{{{{(1 - p)}^2}}} + \frac{{6{p^3}}}{{{{(1 - p)}^3}}}$}.

To compute ${\bar \Delta _i}$ of device $i$, let us consider the area $A$ (see Fig. \ref{fig:aoi_example}) between two consecutive successful updates. Suppose that the last successful update occurs at $t_j^{g + 1}$ and the next successful update occurs at $t_{j + X}^{g + 1}$. $A$ is computed by
\begin{align}	
A = \int_{t_j^{g + 1} = 0}^{t_{j + X}^{g + 1} = XGT} {{\Delta _i}(t)} dt = XGT \cdot T + {(XGT)^2}/2. \label{eqa:area_compute} 
\end{align}
The time slot duration $T$ is related to the block length $L$ and the bandwidth allocated to device $i$, i.e., $T = \frac{{ML}}{B}$. Thus, ${\bar \Delta _i}$ is
\begin{align}	
{\bar \Delta _i} &= \mathop {\lim }\limits_{W \to \infty } \frac{{\sum\limits_{w = 1}^W {{A_w}} }}{{\sum\limits_{w = 1}^W {{X_w}GT} }} = T + \frac{{NTE[{X^2}]}}{{2ME[X]}} \notag \\
&= \left( {1 + \frac{{N(1 + p)}}{{2M(1 - p)}}} \right)\frac{{ML}}{B}. \label{eqa:avg_aoi_final}
\end{align}

\noindent For symmetric channels, the average AoI of the whole network is $\bar \Delta  = \frac{1}{N}\sum\limits_{i = 1}^N {{{\bar \Delta }_i}}  = {\bar \Delta _i}$.

We can see from (\ref{eqa:avg_aoi_final}) that the average AoI of a multichannel access system is affected by the number of subchannels $M$, the block length  $L$, and the decoding successful probability $1-p$. It is worth mentioning that the effective received SNR $\gamma$ in (\ref{eqa:per_bound}) is affected by $M$. For a given received power $P$, the effective received SNR $\gamma$ per user increases monotonically as $M$ increases because the power $P$ from the device is concentrated into a narrower band. As a result, the decoding PER $p$ is also affected by $M$, i.e., $p$ decreases as $M$ increases.

When the effective received SNR $\gamma $ is high enough such that $p \approx 0$, the average AoI becomes
\begin{equation}	\label{eqa:high_snr_avg_aoi}
\bar \Delta  \approx \left( {1 + \frac{N}{{2M}}} \right)\frac{{ML}}{B} = \frac{{2M + N}}{2} \cdot \frac{L}{B}
\end{equation}
Eq. (\ref{eqa:high_snr_avg_aoi}) shows that if the number of users $N$ is fixed, the average AoI is proportional to the number of subchannels $M$.

\subsubsection*{Bounded AoI ${\Delta _{THR}}$} 
We first estimate the bounded AoI ${\Delta _{THR[i]}}$ for device $i$. Specifically, $\overline {\Delta _i^2} $ is computed by
\begin{equation}
\small
\begin{aligned}
\overline {\Delta _i^2}  &= {\lim _{T \to \infty }}\frac{1}{T}\int_0^T {{\Delta _i}{{(t)}^2}} dt = \mathop {\lim }\limits_{W \to \infty } \frac{{\sum\limits_{w = 1}^W {\int_{t_j^{g + 1} = 0}^{t_{j + X}^{g + 1} = {X_w}GT} {{\Delta _i}{{(t)}^2}} dt} }}{{\sum\limits_{w = 1}^W {{X_w}GT} }}\\
&= \frac{{{{N}^2}{L^2}}}{{3{B^2}}} \cdot \frac{{E[{X^3}]}}{{E[X]}} + \frac{{MN{L^2}}}{{{B^2}}} \cdot \frac{{E[{X^2}]}}{{E[X]}} + \frac{{{M^2}{L^2}}}{{{B^2}}} \\
&= \frac{{{L^2}}}{{{B^2}}} \cdot \left[ {\frac{{{N^2}\left( {1 - 8p - 11{p^2} + 24{p^3}} \right)}}{{3{{(1 - p)}^2}}} + \frac{{MN\left( {1 + p} \right)}}{{1 - p}} + {M^2}} \right].
\label{eqa:intermediate_product}
\end{aligned}
\end{equation}
Substituting ${\bar \Delta _i}$ in (\ref{eqa:avg_aoi_final}) and $\overline {\Delta _i^2} $ in (\ref{eqa:intermediate_product}) into (\ref{eqa:upper_bound}), we obtain an upper bound ${\hat \Delta _{THR[i]}}$ of $\Delta _{THR[i]}$, as shown in (\ref{eqa:final_bound_result}).


\begin{equation}
\small
\begin{aligned}
{{\hat \Delta }_{THR[{\rm{i}}]}} &= \sqrt {\frac{{\overline {\Delta _i^2}  - {{\left( {{{\bar \Delta }_i}} \right)}^2}}}{{1 - \epsilon}}}  + {{\bar \Delta }_i} \\
&= \frac{L}{B} \cdot \left( {\sqrt {\frac{{2{N^2} + 3MN - 16{N^2}p - (22{N^2} + 3MN){p^2} + 48{N^2}{p^3}}}{{3{{(1 - p)}^2}(1 - \epsilon)}}}  + \frac{{2M + N + (N - 2M)p}}{{2(1 - p)}}} \right)
\label{eqa:final_bound_result}
\end{aligned}
\end{equation}

With symmetric channels, $\hat \Delta _{THR} = {\hat \Delta _{THR[i]}}$. From (\ref{eqa:final_bound_result}) we see that the bounded AoI is affected by the block length $L$, the bandwidth $B$, the number of subchannels $M$, and the number of IoT devices $N$. For high SNR scenarios with $p \approx 0$, the bounded AoI can be written as
\begin{equation}	\label{eqa:high_snr_bound}
{\hat \Delta _{THR}} \approx \frac{L}{B} \cdot \left( {\sqrt {\frac{{2{N^2} + 3MN}}{{3(1 - \epsilon)}}}  + \frac{{2M + N}}{2}} \right).
\end{equation}
We can see above that when the effective SNR $\gamma$ is high, a larger number of subchannels gives a larger bounded AoI. This is intuitive because a larger number of subchannels leads to longer transmission time. 
However, when $p \neq 0$, it is difficult to establish a clear relationship between the number of subchannels and the bounded AoI from (\ref{eqa:final_bound_result}). We then explore such scenarios by simulations, as detailed in the next section.

\section{Numerical Results}	\label{sec:num_results}
This section presents numerical results of the average AoI and bounded AoI derived in Section \ref{sec:aoi_multichannel}. We show that in an $N$-user information update system, multichannel access with $M$ channels provides a good balance between average AoI and bounded AoI, compared with conventional TDMA ($M=1$) and FDMA systems ($M=N$). 

We assume $N=20$ devices and vary the number of subchannels $M$ from $1$ to $20$. Each source packet has $100$ bits (i.e., $K=100$). We normalize both the system bandwidth and the noise power to $1$ so that the received power $P$ is in units of $dB$ and $P$ is varied from $0dB$ to $8dB$. For each $P$, we optimize the block length of coded bits $L$, ranging from $100$ bits to $400$ bits to obtain the optimal block length for average AoI $\bar \Delta$ and bounded AoI $\hat \Delta _{THR}$, respectively. The PERs associated with different $P$ and $L$ are computed using (\ref{eqa:per_bound}).

\begin{figure}
\centering
\includegraphics[width=\textwidth]{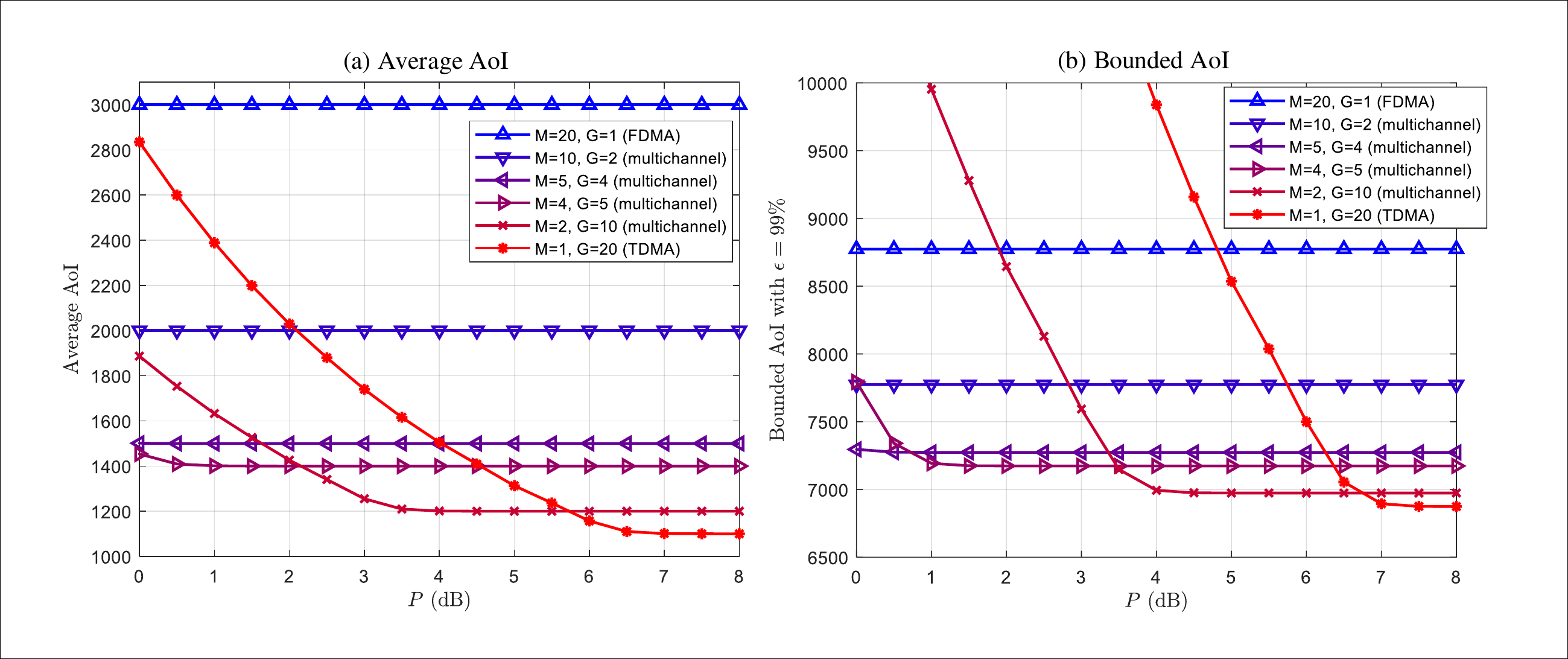}
	\caption{(a) Average AoI and (b) Bounded AoI of the multichannel systems, the TDMA system, and the FDMA system.}\label{fig:plot_aoi_compressed}\label{fig:avg_aoi_plot}\label{fig:bounded_aoi_plot}
\vspace{-0.2in}
\end{figure}

\subsection{Average AoI}
We now take a look at the simulation results of the average AoI $\bar \Delta $ under different numbers of subchannels $M$ and received powers $P$, as shown in Fig. \ref{fig:avg_aoi_plot}. First, we observe that the FDMA system with $M=N=20$ has a very stable but high average AoI. A device in the FDMA system has very high effective received SNR $NP$ (due to a low bandwidth), and consequently the PER is negligible even when $P$ is small, giving a very stable performance across different $P$. On the other hand, the packet transmission time is long due to the narrow bandwidth, thereby increasing the average AoI. Second, we observe that a device in the TDMA system ($M=1$) enjoys the shortest packet transmission time but has the lowest effective SNR $P$. As a result, the average AoI depends much on $P$: when $P$ is low so that the PER is high, a user needs to wait for a long time for the next update opportunity if an update packet is lost, thereby leading to high average AoI; when $P$ increases, the average AoI decreases quickly.

A multichannel access system provides a good balance of average AoI between TDMA and FDMA systems, e.g., when $M$ is small (large), the average AoI performance is closer to that of TDMA (FDMA). In particular, when $P$ is low, (e.g., $P<2.1dB$), the multichannel system with four subchannels ($M=4$) has a lower average AoI than the TDMA system does. More importantly, when $M=4$, the average AoI is stable across different $P$ like an FDMA system; it is also much lower than that of the FDMA system. This shows that by choosing an appropriate $M$, a multichannel system can enjoy the advantages of both TDMA and FDMA systems. 


\subsection{Bounded AoI}
Next, we investigate the effects of the number of subchannels $M$ on bounded AoI, as shown in Fig. \ref{fig:bounded_aoi_plot}. We fix the $\epsilon = 99\% $ for the bounded AoI.

As shown in Fig. \ref{fig:bounded_aoi_plot}, as with average AoI, FDMA gives a more uniform bounded AoI, while TDMA has varying bounded AoI under different $P$. By contrast, the multichannel system can give a low and uniform bounded AoI across different $P$. Specifically, with $M=4$, the multichannel system outperforms the FDMA system under all $P$ due to shorter packet transmission time (although with lower effective SNR). When $P<3.5dB$, the multichannel system has a lower bounded AoI than the TDMA system does due to its higher effective SNR. Although TDMA still has the lowest bounded AoI when $P>7dB$, the multichannel system has a more stable bounded AoI across different $P$. This also indicates multichannel systems are more robust to varying SNRs across different devices by providing low and uniform bounded AoI at the same time.

\begin{figure}
\centering
\includegraphics[width=0.6\textwidth]{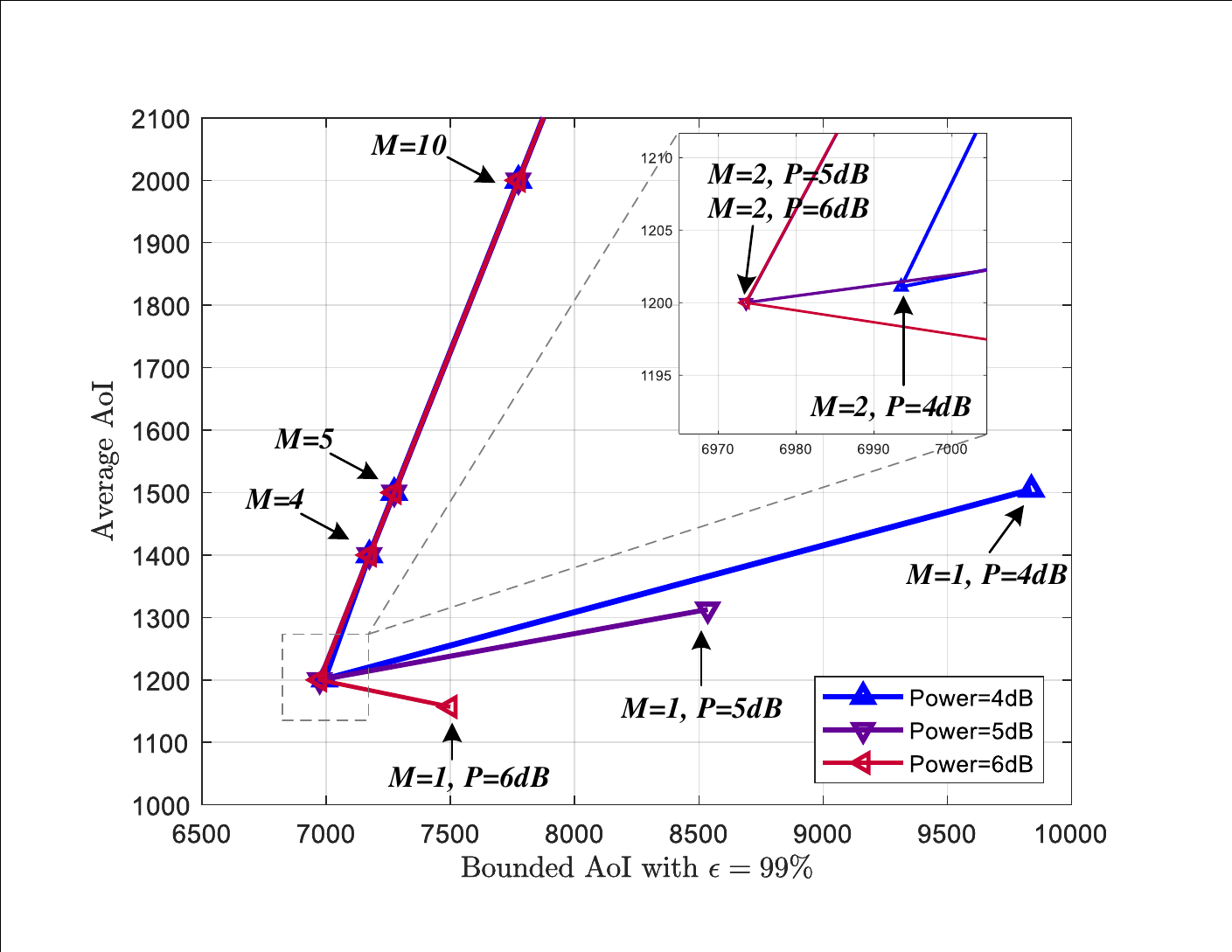}
\caption{Average AoI versus Bounded AoI with $P = 4\sim6dB$.}\label{fig:balance_snr}
\vspace{-0.17in}
\end{figure}

The results above also indicate that when $P$ is high (i.e., $p \approx 0$ when $P \ge 7dB$), TDMA (i.e., $M = 1$) is the optimal setting to obtain both the optimal average AoI and the optimal bounded AoI. However, for the regime of medium and low received powers ($P \le 6dB$), the multichannel system can provide a good balance between average AoI and bounded AoI, as will be shown next. 

\subsection{Balancing between average AoI and bounded AoI}
\begin{figure}
\centering
\includegraphics[width=0.6\textwidth]{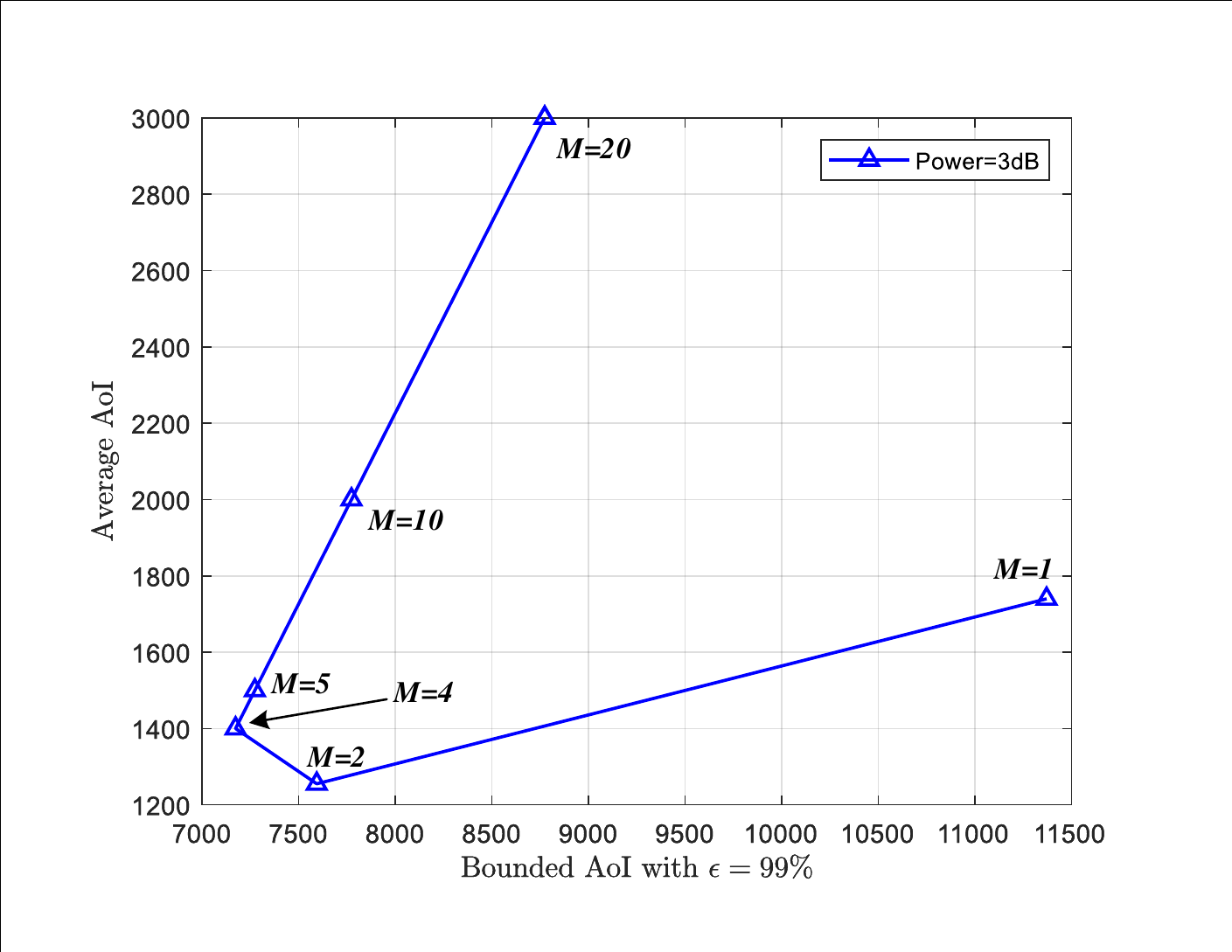}
\caption{Average AoI versus Bounded AoI with $P=3dB$.}\label{fig:mid_snr_balance}
\vspace{-0.12in}
\end{figure}

The interplay between the number of channels $M$ and the received power $P$ on the average AoI $\bar \Delta$ and bounded AoI $\hat \Delta _{THR}$ is shown in Figs. \ref{fig:balance_snr} and \ref{fig:mid_snr_balance}. We plot the average AoI and the bounded AoI that a multichannel system can achieve for a fixed received power $P$. Specifically, for \textit{\underline{each curve}}, we fix $P$ to compute the average AoI $\bar \Delta$ and bounded AoI $\hat \Delta _{THR}$ for $M \in \left\{ {1,2,4,5,10,20} \right\}$ in the simulation. We let $\bar \Delta$ be the x-axis, $\hat \Delta _{THR}$ be the y-axis, and the points ${(\bar \Delta ,{\hat \Delta _{THR}})_M}$ belonging to the same power $P$ are connected to construct a contour. The curves in both figures serve as references to balance the average AoI and the bounded AoI performances under different $P$.

As shown in Fig. \ref{fig:balance_snr}, when the received power is in the mid-range (e.g. $P = 4\;{\rm{or}}\;5dB$), the multichannel system can achieve its best performance from the perspective of both average AoI and bounded AoI when $M=2$. When the power increases to $P = 6dB$, although a lower average AoI is achieved if $M$ is set to $1$, the multichannel system with $M=2$ can still achieve a good balance between the two AoI metrics.

Fig. \ref{fig:mid_snr_balance} shows that when $P=3dB$, going from $M=1$ to $M=2$ or $M=4$, we can achieve substantial improvement in both metrics. The optimal average AoI is achieved when $M=2$ while the optimal bounded AoI is achieved when $M=4$.

\section{Conclusion}	\label{sec:conclusion}
We have investigated multichannel access in terms of information freshness. Specifically, we derive the average AoI under different numbers of subchannels $M$. Moreover, we examine bounded AoI via the estimations of its upper bound. Simulation results show that when all users have the same received power, multichannel access provides low average AoI and uniform bounded AoI simultaneously. 

When $M$ is large (e.g., a pure FDMA system), information update suffers from long transmission time due to small allocated bandwidth, leading to high average AoI. However, a large $M$ gives a more uniform bounded AoI across different received powers. By contrast, a small $M$ leads to high bounded AoI (e.g., a pure TDMA system), which is not suitable for time-critical systems. Multichannel access systems can achieve the best performances from the perspective of both average AoI and bounded AoI. Moreover, our results also indicate that in practical scenarios where users have different received powers, multichannel access is more robust to the varying wireless channels across the users, since it can simultaneously give low average AoI and uniform bounded AoI.


%


\ifCLASSOPTIONcaptionsoff
  \newpage
\fi

\bibliographystyle{IEEEtran}
\bibliography{main}

\end{document}